\documentclass[journal,10pt]{IEEEtran}
\usepackage{times,amsmath,epsfig}
 \usepackage{algorithm}
\usepackage{algorithmic}
\usepackage{enumitem}
\usepackage{chemarrow}
\usepackage{graphicx}
\usepackage{stfloats}
\usepackage{cite}
\usepackage{color}
\usepackage{setspace}
\usepackage{psfrag}
\usepackage{subcaption}
\usepackage{caption}
 \usepackage{amssymb}
 \usepackage{epsfig}
 \usepackage{pifont}
 \usepackage{amsmath}
 \allowdisplaybreaks[4]
 \usepackage{flushend}
 \usepackage{array}
 \usepackage{multicol}
 \usepackage{amsfonts}
 \usepackage{lipsum} 
 \usepackage{multirow}
 \usepackage{verbatim}
 \usepackage{rotating}
\usepackage[acronym]{glossaries}
\usepackage{lscape}
 \usepackage{caption}
 \usepackage{makecell}
 \usepackage[color]{changebar}
 \cbcolor{red}
 \usepackage[switch]{lineno}
 \usepackage{mathtools}

\begin{document}
\title{\LARGE{Distributed Detection and Bandwidth Allocation with Hybrid Quantized and Full-Precision Observations over Multiplicative Fading Channels}}

\author{{Linlin Mao}, {\em Member,~IEEE}, {Zeping Sui}, {\em Member,~IEEE}, \\{Michail Matthaiou}, {\em Fellow,~IEEE}, and {Hongbin Li}, {\em Fellow,~IEEE}
\vspace{-2.5em}
\thanks{This work was supported in part by the National Natural Science Foundation of China under Grants 62371447 and 62192711. (\emph{Corresponding author: Zeping Sui})}
\thanks{Linlin Mao is with the Institute of Acoustics, Chinese Academy of Sciences, Beijing 100190, China. (e-mail: maoll@mail.ioa.ac.cn).}
\thanks{Zeping Sui is with the School of Computer Science and Electronics Engineering, University of Essex, Colchester CO4 3SQ, U.K. (e-mail: zepingsui@outlook.com).}
\thanks{Michail Matthaiou is with the Centre for Wireless Innovation (CWI), Queen’s University Belfast, Belfast BT3 9DT, U.K. (e-mail: m.matthaiou@qub.ac.uk).}
\thanks{Hongbin Li is with the Department of Electrical and Computer Engineering, Stevens Institute of Technology, Hoboken, NJ 07030 USA (e-mail: hongbin.li@stevens.edu).}
}
\maketitle

\begin{abstract}
A hybrid detector that fuses both  quantized and full-precision observations is proposed for weak signal detection under additive and multiplicative Gaussian noise.  We first derive a locally most powerful test (LMPT)--based hybrid detector from the composite probability distribution of the compound observations received by the fusion center, and then analyze its asymptotic detection performance. Subsequently, we optimize the sensor-wise quantization thresholds to achieve near-optimal asymptotic performance at the local sensor level. Moreover, we propose a mixed-integer linear programming approach to solve the optimization problem of transmission bandwidth allocation accounting for bandwidth constraints and error-prone channels. Finally, simulation results demonstrate the superiority of the proposed hybrid detector and the bandwidth allocation strategy, especially in challenging error-prone channel conditions. 
\end{abstract}
\IEEEpeerreviewmaketitle
\vspace{-0.8em}
\section{Introduction}\label{Section1}
Distributed detection in sensor networks has emerged as a pivotal research area with applications spanning various fields, including environmental monitoring, healthcare, and industrial automation~\cite{tabella2023time}. Distributed detection, especially in wireless sensor networks (WSNs), faces challenges, such as constrained energy and limited bandwidth. Prior solutions to mitigate these constraints have considered one-bit quantization of raw observations~\cite{fang2013one} and physical quantities indicative of node information, such as likelihood ratios~\cite{Mohammadi2022generalized}. Although these approaches reduce the data transmission volume and enhance the energy efficiency, they significantly compromise the fidelity of node information and the performance of system detection~\cite{Mao2024Multi}. 

Recently, communication technology advancements have led to increased wireless transmission rates, and the development of energy harvesting techniques has partially alleviated energy scarcity issues in sensor networks (SNs) \cite{meng2024queuing}. Based on these technological strides, efficient distributed detection methods in WSN based on multi-bit quantization have been proposed \cite{yang2023weak,quan2023distributed,lu2023distributed}.  These methods focus on quantizing the data with a predefined bit depth. In practice, the transmission bandwidth of individual nodes in SNs can vary significantly depending on  physical fields, frequency bands, modulation schemes, and energy management strategies \cite{liu2022diffusion,li2023joint}.  Hybrid quantized signal detection that fuses low-bit quantized data with varying quantization levels was investigated in \cite{cheng2019multibit} and \cite{yang2023hybrid}.  While previous studies have explored hybrid quantization, they have not tackled the critical challenge of bandwidth allocation.

Distributed detection of weak signals from one-bit measurements embedded in multiplicative noise was investigated in \cite{wang2019spl}, where a one-bit locally most powerful test (LMPT) detector, assuming error-free reporting channels, was proposed. One-bit distributed detection of a non-cooperative target with a spatial signature was proposed in \cite{ciuonzo2021distributed}, where multiplicative fading and error-prone transmission have been taken into account. Assuming error-free transmission, optimal bit allocation was considered in \cite{luo2019optimal} for target tracking in underwater WSNs with additive and multiplicative noise. To the best of our knowledge, no prior work has yet addressed multi-bit distributed detection in conjunction with its bandwidth allocation problem in multiplicative fading channels.

Against this background, we hereafter design a hybrid detector and a bandwidth allocation algorithm for weak signal detection in combined additive and multiplicative Gaussian noise. Table~\ref{table1} summarizes our contributions, which are described explicitly below:
\begin{itemize}
	\item A hybrid detector that integrates both quantized and full-precision observations in bandwidth-constrained distributed SNs is proposed, where multiplicative fading and error-prone transmission are considered. 
	\item Node-level quantization thresholds are optimized for both error-free and error-prone reporting channels to ensure near-optimal asymptotic performance.
	\item To enhance the overall detection performance at the network level, transmission bandwidth allocation among nodes is optimized by considering the disparity of the error-prone channels.
\end{itemize}
\begin{table}[t]
\small
\centering
\caption{Contrasting Our Work with Prior Research}
\label{table1}
\begin{tabular}{l|c|c|c|c|c}
\hline
Contributions & \textbf{Our work} & \cite{cheng2019multibit} & \cite{yang2023hybrid} & \cite{wang2019spl} & \cite{ciuonzo2021distributed}\\
\hline
\hline
Hybrid quantization & \checkmark & \checkmark & \checkmark &  &  \\
\hline
Error-prone channels & \checkmark & \checkmark &  &  & \checkmark \\
\hline
Multiplicative fading & \checkmark &  &  & \checkmark & \checkmark \\
\hline
Bandwidth allocation & \checkmark &  &  &  &  \\
\hline
\end{tabular}
\vspace{-2em}
\end{table}

The rest of our paper is organized as follows: Section \ref{Section2} introduces the system model for hybrid detection in the presence of both additive and multiplicative Gaussian noise, while Section \ref{Section3} lays out the derivation of the proposed detector. In Section \ref{Section4}, we determine the quantization thresholds at each node level and optimize the bandwidth allocation at the network level. Section \ref{Section5} presents the simulation results, followed by the concluding remarks in Section \ref{Section6}.

\textit{Notations:} Boldface lowercase (uppercase) letters denote vectors (matrices).  $\mathbb{R}^{M \times N}$ and $\mathbb{N}$ are the real matrix and natural number spaces. $(\cdot)^{\mathrm{T}}$, $\mathbb{E}[\cdot]$, $\mathbb{I}(\cdot)$, and $\mathrm{diag}(\cdot)$ denote transposition, expectation, indicator function, and diagonalization, respectively. $\odot$ and $\otimes$ are element-wise and Kronecker products. $[K] \triangleq \{1,\ldots,K\}$ denotes the integer index set from 1 to $K$. $\mathbf{u}_K$ is the all-one vector of dimension $K$, and $\mathbf{e}_K^k$ is the $k$th standard basis vector in $\mathbb{R}^K$. $\sim$ and $\stackrel{\text{a}}{\sim}$ stand for ``distributed as'' and ``asymptotically distributed as'', respectively. $\mathcal{N}(\mu,\sigma^2)$ denotes a Gaussian distribution with mean $\mu$ and variance $\sigma^2$. $P(\cdot)$ and $P(\cdot|\cdot)$ are the probability and conditional probability mass functions.

\vspace{-0.4em}
\section{System Model}\label{Section2}

Consider a distributed SN that employs $M$ geographically dispersed sensors to simultaneously observe a  phenomenon of interest. The objective is to detect an unknown deterministic weak signal $\theta$ amidst multiplicative fading and additive noise:
\begin{align}\label{Eq1}
    \left\{
    \begin{array}{l}
      \mathcal{H}_0:y_m=w_m, \\
      \mathcal{H}_1:y_m=h_m\theta+w_m, \, m=1,2,\ldots, M,
      \end{array}\right.
\end{align}
where $y_m$ denotes the measurements collected by the $m$th sensor; $w_m\sim\mathcal{N}(0,\sigma_n^2)$ and $h_m\sim\mathcal{N}(1,\sigma_h^2)$ represent the additive noise and the multiplicative fading, respectively, which are assumed to be mutually independent across sensors and independent of each other. The nonzero mean assumption  for $h_m$ , widely adopted in studies of wireless sensor networks and distributed detection systems~\cite{zhu2015parameter,wang2019spl}, reflects practical propagation environments with coexisting line-of-sight and multipath components, which are frequently encountered in urban deployments~\cite{gapeyenko2021line} and vehicular communications~\cite{maaref2021autonomous}.

To address channel fading and energy constraints, we partition $M$ sensors into two groups: $M_q$ low-bit sensors and $M_u \triangleq M - M_q$ full-precision ones. The output of the $m$-th low-bit sensor, using a $q_m$-bit quantizer, is given by:
\begin{align}\label{Eq2}
	\pmb{b}_m=
	\left\{
	\begin{array}{ll}
	\pmb{z}_{m,1}\,, &-\infty<y_m<\tau_{m,1},\\
	\pmb{z}_{m,2}\,, &\tau_{m,1}\leq y_m<\tau_{m,2},\\
	\vdots                   &\vdots\\
	\pmb{z}_{m,2^{q_m}}\,, &\tau_{m,2^{q_m}-1}\leq y_m<+\infty,
	\end{array}\right.
\end{align}
where $\left\lbrace \tau_{m,i}\right\rbrace _{i=1}^{2^{q_m}-1}$ denote the quantization thresholds, while $\pmb{z}_{m,i}\triangleq\left[z_{m,i,q_m},z_{m,i,q_m-1},\cdots,z_{m,i,1}\right]^T$ with $z_{m,i,k}\in\{0,1\}$. The codeword $\pmb{b}_m$ is transmitted to the fusion center (FC) over a potentially error-prone wireless link, modeled as a binary symmetric channel (BSC). The probability of $\pmb{z}_{m,j}$ being received as $\pmb{z}_{m,i}$ is given by:
\begin{align}\label{Eq3}
	 P(\pmb{v}_{m}=\pmb{z}_{m,i}|\pmb{b}_{m}=\pmb{z}_{m,j})=\,&P_{\text{e},m}^{D_{m,i,j}}(1-P_{\text{e},m})^{q_m-D_{m,i,j}}\nonumber\\[6pt]
	\triangleq\,&G(q_m,P_{\text{e},m},D_{m,i,j}),
\end{align}
Here, $P_{\text{e},m}$ denotes the crossover probability, and $D_{m,i,j}$ is the Hamming distance between $\pmb{z}_{m,j}$ and $\pmb{z}_{m,i}$.  Due to varying channel conditions, all sensors experience heterogeneous error rates. Low-bit sensors are susceptible to transmission errors, whereas full-precision sensors are assumed to operate over reliable channels with negligible errors. This reliability is underpinned by scheduled wireless access~\cite{zhang2022reliable, wei2021reliable}, robust forward error correction~\cite{wei2021reliable,nasseri2021globally}, or infrastructure support, such as fiber or wired backhaul~\cite{wei2021reliable,madapatha2020integrated}.

Let the data transmitted to the FC by the low-bit sensors be denoted as $\pmb{V}=\left\lbrace \pmb{v}_{1},\pmb{v}_{2},\ldots, \pmb{v}_{M_q}\right\rbrace$ and the data from  the full-precision sensors as $\pmb{\tilde{v}}=\left[\tilde{v}_1,\tilde{v}_2,\ldots, \tilde{v}_{M_u}\right]$, respectively, with $\tilde{v}_{\kappa}\triangleq {y}_{{M_q}+\kappa}, \forall_{\kappa=1}^{M_u} \kappa \in \mathbb{N}$. Note that the analog messages collected by the full-precision nodes are converted into floating-point numbers with bit length  $l_0$ to facilitate communication and reduce the storage requirements.  In this model, bandwidth allocation is assumed to be directly proportional to the number of bits assigned to each sensor, which is a reasonable simplification in digital communication systems~\cite{yang2023hybrid}. Therefore, assigning quantization bits to each sensor is equivalent to allocating bandwidth.  The objective is to optimally configure the low-bit   and full-precision nodes under a total data transmission limit of $Q$ for both the quantized and full-precision messages, as well as  to design the corresponding quantizers and detectors.
\vspace{-0.5em}
\section{The Proposed Hybrid Detector}\label{Section3}
In this section, we design the hybrid detector that fuse the observations from both the low-bit quantization and the full-precision sensors. Then, we analyze the asymptotic detection performance of the proposed detector.
\vspace{-0.7em}
\subsection{The LMPT-based Hybrid Detector}\label{Section3-1}
The detection problem can be recast as a one-sided hypothesis test, with $\mathcal{H}_0: \theta=0$ and $\mathcal{H}_1: \theta\rightarrow 0^+$.  Accordingly, we adopt the LMPT, which is well-suited for detecting weak signals characterized by small mean shifts~\cite{Mao2024Multi,wang2019spl}. The corresponding test statistic is given by \cite{kay1998fundamentals}:
\begin{equation}\label{Eq4}
T_\text{LMPT}=\left(\frac{\partial\ln p(\pmb{V},\pmb{\tilde{v}}|\mathcal{H}_1;\theta)}{\partial \theta}\bigg/\sqrt{\text{FI}(\theta)}\right)_{\theta=0}\underset{H_0}{\mathop{\overset{H_1}{\mathop{\gtrless}}\,}}\eta,
\end{equation}
where $\text{FI}(\theta)$ denotes the Fisher information, which is a scalar independent of the measurements and thus eliminable. Nevertheless, we keep it in \eqref{Eq4} as the scaled test variable possesses a simple asymptotic distribution, as shown in Section III-C. Also, $p(\pmb{V},\pmb{\tilde{v}}|\mathcal{H}_1;\theta)$ denotes the composite probability distribution of the received data, which can be formulated as
\begin{align}\label{Eq5}
	&p(\pmb{V},\pmb{\tilde{v}}|\mathcal{H}_1;\theta)\nonumber\\
	=&\prod_{m=1}^{M_q}\prod_{i=1}^{2^{q_m}}\left[\sum_{j=1}^{2^{q_m}}G(q_m,P_{\text{e},m},D_{m,i,j})Q_{m,j}(\theta)\right]^{I(\pmb{v}_m=\pmb{z}_{m,i})}\nonumber\\
	\times& \prod_{\kappa=1}^{M_u}\frac{1}{\sqrt{2\pi}\sigma_s(\theta,\sigma_h^2,\sigma_n^2)}\exp\left\lbrace-\frac{(\tilde{v}_{\kappa}-\theta)^2}{2\sigma_s^2(\theta,\sigma_h^2,\sigma_n^2)}\right\rbrace
\end{align}
where 
\begin{align}\label{Eq6}
Q_{m,j}(\theta)\triangleq \Phi\left(\frac{\tau_{m,j-1}-\theta}{\sigma_m(\theta,\sigma_h^2,\sigma_n^2)\!}\right)-\Phi\left(\frac{\tau_{m,j}-\theta}{\sigma_m(\theta,\sigma_h^2,\sigma_n^2)}\right)
\end{align}
with $\Phi(x)\triangleq1/\sqrt{2\pi}\int_{x}^{+\infty}\exp(-\alpha^2/2)\,\text{d}\alpha$ and $\sigma_s^2(\theta,\sigma_h^2,\sigma_n^2)\allowbreak=\theta^2\sigma_h^2+\sigma_n^2$. Taking the derivative of the logarithm of \eqref{Eq5} with respect to $\theta$ leads to 
\begin{align}\label{Eq7}
&\frac{\partial\ln p(\pmb{V},\pmb{\tilde{v}}|\mathcal{H}_1;\theta)}{\partial \theta}\nonumber\\
=&\sum_{m=1}^{M_q}\sum_{i=1}^{2^{q_m}}\!\left[\!\frac{I(\pmb{v}_m=\pmb{z}_{m,i})}{\sigma_s^3(\theta,\sigma_h^2,\sigma_n^2)}\frac{\!\sum_{j=1}^{2^{q_m}}\!G(q_m,P_{\text{e},m},D_{m,i,j})F_{m,j}(\theta)}{\sum_{j=1}^{2^{q_m}}\!G(q_m,P_{\text{e},m},D_{m,i,j})Q_{m,j}(\theta)}\right]\nonumber\\
+&\sum_{\kappa=1}^{M_u}\left[\frac{\tilde{v}_{\kappa}-\theta-\theta\sigma_h^2}{\sigma_s^2(\theta,\sigma_h^2,\sigma_n^2)}+\frac{\theta\sigma_h^2(\tilde{v}_{\kappa}-\theta)^2}{\sigma_s^4(\theta,\sigma_h^2,\sigma_n^2)}\right],
\end{align}
where 
\begin{align}\label{Eq8}
F_{m,j}(\theta)&=\left(\sigma_n^2+\theta\tau_{m,j-1}\sigma_h^2\right)\Psi\left({\tau_{m,j-1}}/\textstyle{\sigma_s(\theta,\sigma_h^2,\sigma_n^2)}\right)\nonumber\\
&-\left(\sigma_n^2+\theta\tau_{m,j}\sigma_h^2\right)\Psi\left({\tau_{m,j}}/\textstyle{\sigma_s(\theta,\sigma_h^2,\sigma_n^2)}\right).
\end{align}
with $\Psi(x)=1/\sqrt{2\pi}\exp(-x^2/2)$. Accordingly, the Fisher information can be formulated as 
\begin{align}\label{Eq9}
\!\text{FI}(\theta)
&\!\triangleq\!-\mathbb{E}\left[\!\frac{\partial^2\ln p(\pmb{V},\pmb{\tilde{v}}|\mathcal{H}_1;\theta)}{\partial \theta^2}\right]\nonumber\\
&\!=\!\frac{1}{\sigma_s^3(\theta,\!\sigma_h^2,\!\sigma_n^2)\!}\!\sum_{m=1}^{M_q}\!\sum_{i=1}^{2^{q_m}}\!\frac{\left[\!\sum_{j=1}^{2^{q_m}}\!G(q_m,\!P_{\text{e},m},\!D_{m,i,j}\!)\!F_{m,j}\!(\theta)\!\right]^2\!}{\sum_{j=1}^{2^{q_m}}\!G(q_m,P_{\text{e},m},\!D_{m,i,j})Q_{m,j}(\theta)}\nonumber\\
&+\frac{M_u}{\sigma_s^2(\theta,\sigma_h^2,\sigma_n^2)}\left[1+\frac{2\theta^2\sigma_h^4}{\theta^2\sigma_h^2+\sigma_n^2}\right].
\end{align}
By substituting $\theta=0$ into \eqref{Eq7} and \eqref{Eq9}, the LMPT detector based on hybrid observations is given by 
\begin{align}\label{Eq10}
\hspace*{-0.75em}T_\text{LMPT\!}\!&\propto\!\sigma_n^{-3}\!\sum_{m=1}^{M_q}\sum_{i=1}^{2^{q_m}}I(\pmb{v}_m=\pmb{z}_{m,i})\quad\quad\quad\nonumber\\
&\times\! \frac{\!\sum_{j=1}^{2^{q_m}}G(q_m,P_{\text{e},m},D_{m,i,j})F_{m,j}(0)}{\!\sum_{j=1}^{2^{q_m}}G(q_m,P_{\text{e},m},D_{m,i,j})Q_{m,j}(0)}
\!\!+\!\sigma_n^{-3}\!\sum_{\kappa=1}^{M_u}\tilde{v}_{\kappa}.
\end{align}
\subsection{Asymptotic Detection Performance}\label{Section3-3}
According to \cite{kay1998fundamentals}, the asymptotic distribution of the LMPT test statistic $T_\text{LMPT}$ in \eqref{Eq10} can be derived as
\begin{align}\label{Eq11}
T_\text{LMPT}\stackrel a\sim\left\{
\begin{array}{ll}
\mathcal{N}(0,1), &\text{under}\quad \mathcal{H}_0\\
\mathcal{N}(\lambda,1), &\text{under}\quad \mathcal{H}_1,
\end{array}\right.
\end{align}
where $\lambda=\theta\sqrt{\text{FI}(0)}$ represents the non-centrality parameter.  The asymptotic behavior is considered as the number of sensors, $M$, tends to infinity. According to the Central Limit Theorem, as $M$ increases, the distribution of $T_\text{LMPT}$ asymptotically approaches a normal distribution~\cite{kay1998fundamentals}. Consequently, given a specified threshold $\eta$, the probability of false alarm can be formulated as
\begin{align}\label{Eq12}
	P_\text{FA}=P(T_\text{LMPT}>\eta|\mathcal{H}_0)\approx\Phi(\eta).
\end{align} 
Similarly, the probability of detection can be given as
\begin{align}\label{Eq13}
	P_\text{D}=P(T_\text{LMPT}>\eta|\mathcal{H}_1)\approx\Phi_{\lambda}(\eta),
\end{align}
where $\Phi_{\lambda}(\beta)=1/\sqrt{2\pi}\int_{\beta}^{+\infty}\exp(-(\alpha-\lambda)^2/2)\text{d}\alpha$ denotes the complementary
cumulative density function for a non-central normal distribution with non-centrality parameter $\lambda$. 
\section{Quantizer Design and Bandwidth Allocation Optimization}\label{Section4}
In this section, we first determine the  quantization thresholds for the low-bit nodes to ensure near-optimal asymptotic performance. Then, we optimize the allocation of transmission bandwidth among nodes within the data transmission limit $Q$ to enhance the overall detection performance. 
\subsection{Low-Bit Quantizer Design}\label{Section4-1}
As shown in \eqref{Eq12} and \eqref{Eq13}, the detection performance improves as $\lambda$ increases, which correlates positively with $\text{FI}(0)$. Therefore, the low-bit quantizer design can be formulated as an optimization problem based on $\text{FI}(0)$, yielding,
\begin{align}\label{Eq14}
	\!\underset{\{\!\pmb{\tau}_m\}_{m=1}^{M_q}}{\!\max}\!\sum_{m=1}^{M_q}\sum_{i=1}^{2^{q_m}}\frac{\big[\sum_{j=1}^{2^{q_m}}\!G(q_m,\!P_{\text{e},m},\!D_{m,i,j})F_{m,j}(0)\big]^2}{\sum_{j=1}^{2^{q_m}}\!G(q_m,\!P_{\text{e},m},\!D_{m,i,j})Q_{m,j}(0)}.\!
\end{align}
Assuming independence across the reporting channels from the low-bit quantized sensors to the FC, \eqref{Eq14} can be decomposed  into $M_q$ separate  sub-problems as
\begin{align}\label{Eq15}
\underset{\{\!\pmb{\tau}_m\!\}_{m=\!1}^{M_q}}{\!\max}\sum_{i=1}^{2^{q_m}}\frac{\big[\sum_{j=1}^{2^{q_m}}\!G(q_m,\!P_{\text{e},m},\!D_{m,i,j})F_{m,j}(0)\big]^2}{\sum_{j=1}^{2^{q_m}}\!G(q_m,\!P_{\text{e},m},\!D_{m,i,j})Q_{m,j}(0)}\!\nonumber\\
\text{s.t.\quad} -\infty<\tau_{m,1}<\cdots<\tau_{m,2^{q_m}-1}<+\infty.
\end{align}

\begin{figure}[t]
\centering
\includegraphics[width=\linewidth]{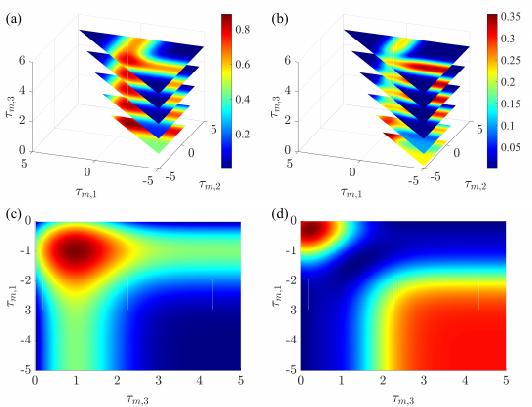}
\caption{3D slice of $\text{FI}(0)$ along \(\tau_{m,3}\) with 2-bit quantization when \(\sigma_n^2 = 1\): (a)  \(P_{\text{e},m}=0\)  and (c) its 2D top-view heatmap ($\text{FI}(0)$ vs \(\tau_{m,1}\), \(\tau_{m,3}\)  with \(\tau_{m,2}=0\) ); (b) \(P_{\text{e},m}=0.2\) and (d) the corresponding 2D top-view heatmap.}
\label{Figure1}
\end{figure}

 For the special case $P_{\text{e},m}=0, \forall m \in [M_q]$, the objective function  to be optimized in \eqref{Eq15} simplifies to
\begin{align}\label{Eq16}
\Xi(\pmb{\tau}_m)=\sum_{i=1}^{2^{q_m}}\frac{F_{m,i}^2(0)}{Q_{m,i}(0)}.
\end{align} 
Define $\triangledown\Xi(\pmb{\tau}_m)=\bigg[\frac{\partial \Xi(\pmb{\tau}_m)}{\partial \tau_{m,1}},\frac{\partial \Xi(\pmb{\tau}_m)}{\partial \tau_{m,2}},\cdots,\frac{\partial \Xi(\pmb{\tau}_m)}{\partial \tau_{m,{2^{q_m}-1}}} \bigg]^T$  as the gradient vector of $\Xi(\pmb{\tau}_m)$ with respect to $\pmb{\tau}_m$,  whose  $i$th element can be formulated as
\begin{align}\label{Eq17}
\frac{\partial \Xi(\pmb{\tau}_m)}{\partial \tau_{m,i}}
&=\frac{\Psi(\frac{\tau_{m,i}}{\sigma_n})}{\sigma_n^3}\frac{F_{m,i}(0)Q_{m,i+1}(0)-F_{m,i+1}(0)Q_{m,i}(0)}{Q_{m,i}(0)Q_{m,i+1}(0)}\nonumber\\
&\times\left[\frac{2\tau_{m,i}}{\sigma_n}-\frac{F_{m,i+1}(0)}{\sigma_n^2Q_{m,i+1}(0)}-\frac{F_{m,i}(0)}{\sigma_n^2Q_{m,i}(0)}\right].
\end{align}
Following the approach in \cite{yang2023hybrid}, we can show that the first term in \eqref{Eq17} is positive, the second is negative, and the third is strictly increasing with $\tau_{m,i}, \forall_{i=1}^{2^{q_m}-1} i\in \mathbb{N}$, which leads us to conclude that the objective function in \eqref{Eq16} is unimodal for all $\tau_{m,i}$ that satisfy the constraint in \eqref{Eq15}.  A three-dimensional slice of $\text{FI}(0)$ along $\tau_{m,3}$ with 2-bit quantization and its 2D top-view heatmap, illustrating $\text{FI}(0)$ values as a function of $\tau_{m,1}$ and $\tau_{m,3}$ with $\tau_{m,2} = 0$, are shown in Figs.\,\ref{Figure1}(a) and (c). When $P_{\text{e},m} = 0$, Fig.\,\ref{Figure1}(c) exhibits a single peak at $\tau_{m,1} = -1$, $\tau_{m,2} = 0$, and $\tau_{m,3} = 1$. In this case, the batch gradient descent algorithm (BGDA) can be used to compute the optimal quantization thresholds for the low-bit quantizers. 

In contrast, when $P_{\text{e},m} \neq 0$, as  shown in Figs.\,\ref{Figure1}(b) and (d), the objective function in \eqref{Eq15} becomes non-unimodal, with two distinct peaks: one at $\tau_{m,1} = -0.2384$, $\tau_{m,2} = 0$, $\tau_{m,3} = 0.2384$ and the other at $\tau_{m,1} = -4.237$, $\tau_{m,2} = 0$, $\tau_{m,3} = 4.237$. This behavior undermines the BGDA's convergence, increasing the risk of slow convergence or getting trapped in local minima. In this case, we resort to a particle swarm optimization approach (PSOA) to solve \eqref{Eq15}  thanks to its well-established effectiveness in optimizing non-unimodal objective functions~\cite{Mao2024Multi,yang2023weak,cheng2019multibit,yang2023hybrid}.
\vspace{-1em}
\subsection{Bandwidth Allocation Optimization}\label{Section4-2}
We now  focus on optimizing the system-level bandwidth allocation under a constrained transmission budget. 
Each sensor is associated with an error probability \(P_{\text{e},m}\), which reflects the heterogeneous link quality across the network. Let \(\pmb{\epsilon} = [\epsilon_1, \epsilon_2, \ldots, \epsilon_N]\) denote the sorted vector of unique error probabilities, and let \(\pmb{f} = [f_1, f_2, \ldots, f_N]\) represent their empirical distribution, where \(f_n\) is the relative frequency of \(\epsilon_n\). Sensors in each error category initially transmit low-bit quantized data and suffer performance degradation due to channel errors. Under sufficient bandwidth, a subset of these sensors can be promoted to full-precision mode and reassigned to reliable communication paths. Let \(a_n\) denote the number of sensors in category \(\epsilon_n\) that are upgraded to full precision and thus assumed to be error-free. 

Let \(L = \max_m q_m\)  be the number of quantization levels. Define the allocation matrix \(\pmb{X} \in \mathbb{N}^{L \times N}\), where \(x_{ln}\) is the number of sensors assigned to quantization level \(l\) and error category \(\epsilon_n\). The corresponding FI values are given by \(\pmb{\Gamma} \in \mathbb{R}^{L \times N}\), with \(\gamma_{ln}\) quantifying the FI contribution of a single low-bit sensor using \(l\) bits under error rate \(\epsilon_n\), as derived from the first term in \eqref{Eq9}. The constant \(\gamma_0\) denotes the per-sensor FI for full-precision transmission, given by the second term. Finally, let \(\pmb{\Lambda} = \mathrm{diag}([L])\) and \(\pmb{d} = [1, 2, \ldots, L]^T\) represent the bit-widths for bandwidth calculation. The bandwidth allocation problem is formulated as:
\begin{align}\label{Eq18}
&\underset{\pmb{X}, \{m_n\}}{\text{max}} 
&& \pmb{u}_L^T \left(\pmb{\Gamma} \odot \pmb{X}\right) \pmb{u}_N + \left( \sum_{n=1}^N a_n \right) \gamma_0 \\
&\text{subject to}
&& \pmb{u}_L^T \pmb{X} \pmb{u}_N +  \sum_{n=1}^N a_n = M, \tag{C1} \label{eq:C1} \\
&&& \pmb{u}_N^T \pmb{X}^\top \pmb{\Lambda} \pmb{u}_L + l_0\sum_{n=1}^N a_n = Q, \tag{C2} \label{eq:C2} \\
&&& \pmb{u}_L^T \pmb{X} \pmb{e}_N^n =  f_n M - a_n, \quad \forall n \in [N], \tag{C3} \label{eq:C3} \\
&&&  0 \leq a_n \leq f_n M,\quad a_n \in \mathbb{N}, \quad \forall n \in [N], \tag{C4} \label{eq:C4} \\
&&& x_{ln} \in \mathbb{N}, \quad  \forall (l,n) \in [L] \times [N]. \tag{C5} \label{eq:C5}
\end{align}
The objective maximizes the total FI contributed by both low-bit and promoted full-precision sensors, while constraints (C1)--(C2) enforce the total sensor count and bandwidth budget. Constraint (C3) preserves per-category allocation after promotion, while (C4)--(C5) impose integrality and feasibility.

To reformulate the problem as a standard integer linear program (ILP)~\cite{junger200950}, define the decision vector  \(\pmb{x} = [\mathrm{vec}(\pmb{X})^T, \pmb{a}^T]^T \in \mathbb{Z}_+^{LN+N}\), where \(\pmb{a} = [a_1,\ldots,a_N]^T\) denotes the number of promoted full-precision sensors per error category. Let \(\pmb{c} = [\mathrm{vec}(\pmb{\Gamma})^T, \gamma_0\,\pmb{u}_N^T]^T\) be the cost vector. The equivalent ILP is:
\begin{align}\label{Eq19}
\min_{\pmb{x} \in \mathbb{Z}_+^{LN+N}} \quad & -\pmb{c}^T \pmb{x} \nonumber \\
\text{s.t.} \quad & \pmb{A}\pmb{x} = \pmb{b}, \pmb{0} \leq \pmb{x} \leq \pmb{u}_b,
\end{align}
where the upper bound vector is given by \(\pmb{u}_{b} = [\infty\,\pmb{u}_{LN}^T, (M\pmb{f})^T]^T\), and the equality constraint system is:
\begin{align}\label{Eq20}
\pmb{A} &= \begin{bmatrix}
\pmb{u}_{LN}^T & \pmb{u}_N^T \\
(\pmb{d} \otimes \pmb{I}_N)^T & l_0\pmb{u}_N^T \\
\pmb{I}_N \otimes \pmb{u}_L^T & \pmb{I}_N
\end{bmatrix}, \quad
\pmb{b} = \begin{bmatrix}
M \\ Q \\ \pmb{f} M
\end{bmatrix}.
\end{align}
The problem can be efficiently solved using MATLAB's \texttt{intlinprog}. A detailed implementation is provided in \textbf{Algorithm~\ref{alg:IP-Bandwidth-Allocation}}.
\begin{algorithm}[t]
\caption{ILP-Based Bandwidth Allocation}
\label{alg:IP-Bandwidth-Allocation}
\begin{algorithmic}[1]
\REQUIRE \(P_{\text{e},1},\ldots,P_{\text{e},M},\ M,\ Q,\ L,\ l_0\)
\ENSURE Bandwidth allocation \(\pmb{X} \in \mathbb{Z}_+^{L \times N}\), promotion vector \(\pmb{a} \in \mathbb{Z}_+^N\)

\item[]  \textbf{Step 1: Error categorization}
\STATE Sort all \(P_{\text{e},m}\), remove duplicates to form \(\pmb{\epsilon} = [\epsilon_1,\ldots,\epsilon_N]\)
\STATE Count occurrences to obtain \(\pmb{f} = [f_1,\ldots,f_N]\); set \(N = |\pmb{\epsilon}|\)

\item[]  \textbf{Step 2: Compute per-sensor Fisher information}
\STATE Evaluate \(\gamma_0\) using~\eqref{Eq9} with \(M_u = 1\)
\FOR{\(l = 1\) to \(L\)}
    \FOR{\(n = 1\) to \(N\)}
        \STATE Compute \(\gamma_{ln}\) from~\eqref{Eq9} under \(q_m = l\), \(P_{\text{e},m} = \epsilon_n\)
    \ENDFOR
\ENDFOR
\STATE Form \(\pmb{\Gamma} \in \mathbb{R}^{L \times N}\)

\item[]  \textbf{Step 3: Construct the objective function}
\STATE Set cost vector \(\pmb{c} = [\mathrm{vec}(\pmb{\Gamma}); \gamma_0\,\pmb{u}_N]\)

\item[]  \textbf{Step 4: Formulate constraints}
\STATE Build equality constraint matrix \(\pmb{A}\) and vector \(\pmb{b}\) from \eqref{Eq20}
\STATE Set upper bounds \(\pmb{u}_b = [\infty\,\pmb{u}_{LN}^T, (M \pmb{f})^T]^T\)

\item[]  \textbf{Step 5: Solve the ILP}
\STATE Solve \(\min -\pmb{c}^T \pmb{x}\) subject to \(\pmb{A}\pmb{x} = \pmb{b},\ 0 \leq \pmb{x} \leq \pmb{u}_b\) via \texttt{intlinprog}

\item[]  \textbf{Step 6: Recover solution}
\STATE Reshape first \( LN \) entries of \( \pmb{x}\) into matrix \( \pmb{X} \in \mathbb{Z}_+^{L \times N} \)
\STATE Extract remaining entries into vector \( \pmb{a} \in \mathbb{Z}_+^N \)

\RETURN \(\pmb{X}, \pmb{a}\)
\end{algorithmic}
\end{algorithm}

Complexity Requirements: The computational complexity of \textbf{Algorithm~\ref{alg:IP-Bandwidth-Allocation}} is dominated by its ILP-solving stage (Step 5). The decision variable vector $\mathbf{x}$ consists of $N(L+1)$ integer variables, each with an approximate uniform upper bound $U \approx M/N + 1$. This structure results in a worst-case complexity of $\mathcal{O}\left( \left( M/N + 1 \right)^{N(L+1)} \right)$. Despite this exponential worst-case complexity, the algorithm remains computationally feasible for small values of $N$ and $L$, ensuring its applicability in the targeted scenarios.

\section{Performance Analysis}\label{Section5}
In this section, simulation results are presented to illustrate the performance of the proposed hybrid detector and the node configuration method.  For comparison, we adopt the following parameter settings throughout this paper:  $\theta=0.25$, $\sigma_n^2=1$, and $\sigma_h^2=0.5$.  For the PSOA parameters, following empirical results and standard PSO configurations for similar optimization tasks~\cite{cheng2019multibit}, we set: acceleration coefficients $c_1 = c_2 = 2.05$, population boundary $\tau_{\text{max}} = 5$, population size $100$, and velocity stopping tolerance $v_{\text{tol}} = 10^{-6}$.

\begin{figure}[htbp]
\centering
\includegraphics[width=\linewidth]{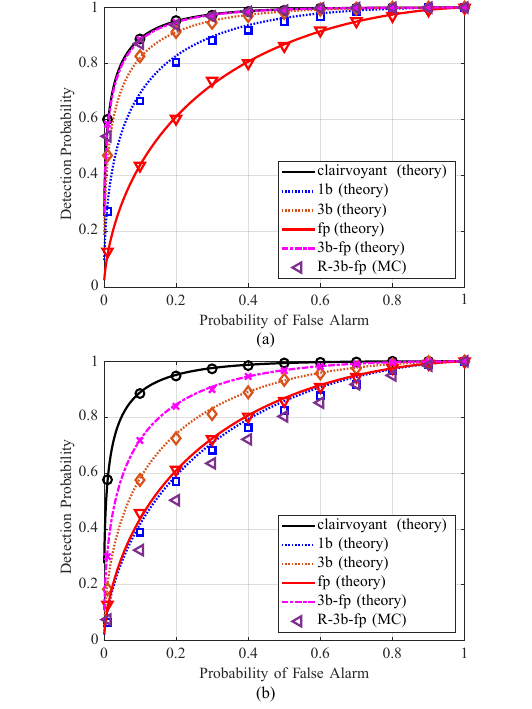}
\caption{ROC performance comparison of multiple detectors, including clairvoyant, quantized (1b, 3b), full-precision (fp), hybrid (3b-fp), and reconstruction-based hybrid (R-3b-fp) detectors, alongside theoretical and MC results, where $M_q=80$, $M_u=20$, $\theta=0.25$, $\sigma_n^2=1$, and $\sigma_h^2=0.5$: (a) $P_\text{e}=0$, (b) $P_\text{e}=0.2$.}
\label{Figure2}
\vspace{-1em}
\end{figure}

We consider a system with $M_q = 80$ quantized sensors (employing either 1-bit or 3-bit quantization) and $M_u = 20$ full-precision (32-bit) sensors. For all quantized sensors, the error probability is uniform, i.e., $P_{\text{e},m}=P_{\text{e}},  \forall m\in [M_q]$. Figure~\ref{Figure2} shows the ROC curves for several detectors: 
the ``clairvoyant'' detector~\cite{wang2019spl,kay1998fundamentals} using ideal analog measurements from all $100$ sensors; 
the ``1b''~\cite{wang2019spl} and ``3b'' detectors using quantized data from $80$ sensors; 
the ``fp'' detector using full-precision data from 20 sensors; 
the ``3b-fp'' detector integrating ``3b'' data with ``fp'' data via Section~\ref{Section3-1}; 
the ``R-3b-fp'' detector reconstructing 3-bit data and averaging it with ``fp'' observations; 
the ``theory'' curve showing asymptotic performance derived from \eqref{Eq13}; 
and the ``MC'' curve from 5,000-trial Monte-Carlo results.
For methods with both theory and MC results, same-colored markers denote MC simulations, while the MC legend is omitted for brevity.

 As shown in Fig.\,\ref{Figure2}(a), under ideal channels, the hybrid detectors 3b-fp and R-3b-fp nearly match the clairvoyant detector, with 3b and 1b following closely and significantly outperforming fp. This demonstrates that low-bit quantizers achieve superior performance with reduced transmission rate (240 bits for 3b, 80 for 1b, versus 640 for fp). Under non-ideal channels ($P_e = 0.2$, Fig.\,\ref{Figure2}(b)), R-3b-fp and 1b performance degrades substantially, whereas the proposed 3b-fp and 3b detectors remain robust due to explicit channel state consideration and the inherent robustness of multi-bit quantization.

\begin{figure}[t]
\centering
\includegraphics[width=\linewidth]{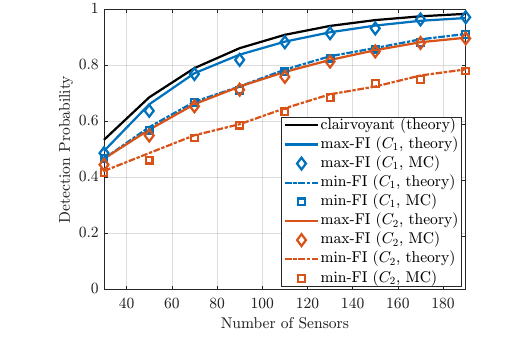}
\caption{Detection probability versus the number of sensors for the proposed hybrid detector with bandwidth allocation optimization, comparing FI-maximizing and FI-minimizing strategies  with parameters: $Q=500$, $\theta=0.25$, $\sigma_n^2=1$, $\sigma_h^2=0.5$, $P_{\text{FA}}=0.1$, $P_{\text{e}} \in \{0, 0.01, 0.1, 0.2\}$, case $C_1$ ($\pmb{f} = [0.6, 0.2, 0.1, 0.1]$), and $C_2$ ($\pmb{f} = [0.1, 0.1, 0.2, 0.6]$).}
\label{Figure3}
\end{figure}

In contrast to the assumption of uniform error probability, we now consider a nonuniform setting with heterogeneous \(P_{\text{e}} \) values to evaluate the performance of our bandwidth allocation strategy. For a fixed bandwidth budget of \(Q=500\), Fig.\,\ref{Figure3} plots the hybrid detector's performance against the number of sensors, comparing two strategies: one that maximizes and the other that minimizes the FI. We consider error probabilities \(P_{\text{e}} \in \{0, 0.01, 0.1, 0.2\}\). Two cases are investigated: Case 1 ($C_1$), where the empirical distribution is \(\pmb{f} = [0.6, 0.2, 0.1, 0.1]\), and Case 2 ($C_2$), where \(\pmb{f} = [0.1, 0.1, 0.2, 0.6]\). The figure illustrates that optimizing the bandwidth allocation using the methodology from Section\,\ref{Section4-2} significantly enhances detection performance. This performance gain is particularly pronounced under more adverse channel conditions.

\begin{figure}[t]
\centering
\includegraphics[width=\linewidth]{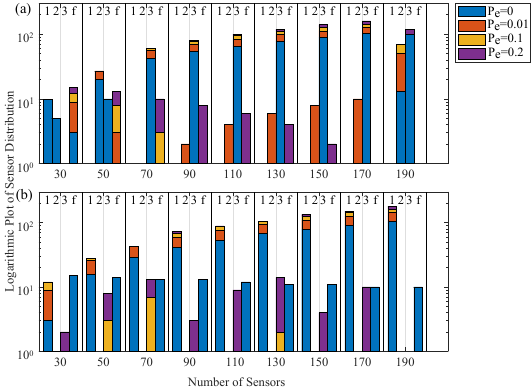}
\caption{Sensor distribution versus the  number of sensors for the proposed hybrid detector with bandwidth allocation optimization that (a) maximizes the FI and (b) minimizes the FI, where $Q=500$, $\theta=0.25$, $\sigma_n^2=1$, $\sigma_h^2=0.5$, $P_\text{FA}=0.1$, $P_{\text{e}} \in \{0, 0.01, 0.1, 0.2\}$, and $\pmb{f} = [0.6, 0.2, 0.1, 0.1]$.}
\vspace{-1.5em}
\label{Figure4}
\end{figure}

For $C_1$, Figures\,\ref{Figure4}(a) and (b) illustrate the sensor distribution corresponding to max-FI and min-FI in Fig.\,\ref{Figure3}, respectively. For each $M$, four bars represent the quantization levels $l = 1, 2, 3$ and full precision (fp), labeled ``1'', ``2'', ``3'', and ``f'', respectively. Each bar is segmented into four error classes, as detailed in the legend. Bars corresponding to zero values remain unfilled. By combining the insights from Fig.\,\ref{Figure4}(a) and Fig.\,\ref{Figure3}, it is evident that within the given bandwidth constraints, a hybrid detector aiming for optimal detection performance should prioritize the deployment of 2b and fp nodes when the SN has a limited number of sensors. As the sensor count in the SN grows, the strategy shifts to reducing the fp nodes while deploying more 2b and 3b nodes, which are more efficient in terms of bandwidth usage and detection capabilities. 
Conversely, Fig.\,\ref{Figure4}(b) illustrates a suboptimal strategy: an excessive allocation to 1-bit and fp nodes. This configuration is inefficient because fp nodes are bandwidth-intensive, leaving insufficient resources for low-bit quantizers. Consequently, the system may be forced to rely on 1-bit quantization to meet the bandwidth constraint. However, while 1-bit nodes are simple, they provide lower FI and are highly susceptible to errors in non-ideal channels.

\vspace{-0.5em}
\section{Summary and Conclusions}\label{Section6}
A LMPT-based hybrid detector that fuses both quantized and full-precision observations was developed for weak signal detection in bandwidth-constrained distributed sensor networks with multiplicative fading. Leveraging the asymptotic distribution of the LMPT test statistic, we optimized the quantization thresholds at each low-bit node to provide near-optimal asymptotic performance at the node level. Specifically, under error-free channels between the FC and nodes, we demonstrated the unimodality of the objective function with respect to each quantization threshold. In such cases, the batch gradient descent algorithm was employed to determine the optimal quantization thresholds. Subsequently, we optimized the allocation of transmission bandwidth among the nodes within the given data transmission limit to enhance overall detection performance. Finally, simulation results demonstrated the superiority of the proposed detector, confirming the importance of hybrid quantization and optimum bandwidth allocation.
\vspace{-1em}
\renewcommand{\refname}{References}
\mbox{} 
\nocite{*}
\bibliographystyle{IEEEtran}

\end{document}